\def\year{2021}\relax
\documentclass[letterpaper]{article} 
\usepackage{aaai21}  
\usepackage{times}  
\usepackage{helvet} 
\usepackage{courier}  
\usepackage[hyphens]{url}  
\usepackage{graphicx} 
\usepackage{amsmath,multirow,booktabs}
\usepackage[switch]{lineno}

\usepackage{natbib}  
\usepackage{caption} 
\title{Phonetic Posteriorgrams based Many-to-Many Singing Voice Conversion via Adversarial Training}
\author{ \Large Haohan Guo\textsuperscript{\rm 1,2}\thanks{Work performed as an intern at Tencent AI Lab. Lei Xie is the corresponding author.}, Heng Lu\textsuperscript{\rm 2}, Na Hu\textsuperscript{\rm 2}, Chunlei Zhang\textsuperscript{\rm 2}, Shan Yang\textsuperscript{\rm 1,2}, Lei Xie\textsuperscript{\rm 1}, Dan Su\textsuperscript{\rm 2}, Dong Yu\textsuperscript{\rm 2}\\
\textsuperscript{\rm 1} School of Computer Science and Engineering, Northwestern Polytechnical University, Xi’an, China\\
\textsuperscript{\rm 2} Tecent AI Lab, Beijing, China\\
hhguo,lxie@nwpu-aslp.org, bearlu@tencent.com
}

\begin{document}
\maketitle
\begin{abstract}

This paper describes an end-to-end adversarial singing voice conversion (EA-SVC) approach. It can directly generate arbitrary singing waveform by given phonetic posteriorgram (PPG) representing content, F0 representing pitch, and speaker embedding representing timbre, respectively. Proposed system is composed of three modules: generator $G$, the audio generation discriminator $D_{A}$, and the feature disentanglement discriminator $D_F$. The generator $G$ encodes the features in parallel and inversely transforms them into the target waveform. In order to make timbre conversion more stable and controllable, speaker embedding is further decomposed to the weighted sum of a group of trainable vectors representing different timbre clusters. Further, to realize more robust and accurate singing conversion, disentanglement discriminator $D_F$ is proposed to remove pitch and timbre related information that remains in the encoded PPG. Finally, a two-stage training is conducted to keep a stable and effective adversarial training process. Subjective evaluation results demonstrate the effectiveness of our proposed methods. Proposed system outperforms conventional cascade approach and the WaveNet based end-to-end approach in terms of both singing quality and singer similarity. Further objective analysis reveals that the model trained with the proposed two-stage training strategy can produce a smoother and sharper formant which leads to higher audio quality. 

\end{abstract}
\section{Introduction}
\label{sec:intro}

Many-to-Many singing voice conversion (SVC) is a technology that converts the timbre of arbitary source song to arbitary target singer's voice. It has been paid great attention due to its potential applications in various interaction and entertainment scenarios. Compared with voice conversion (VC, mainly for speech), the high precision in pitch generation in SVC brings a higher requirement for the model capability to ensure stable and high-quality conversion. To achieve this goal, many algorithms have been tested and proposed. Majority state-of-the-art unsupervised many-to-many SVC approaches are based on autoencoder framework. In the training process, features representing content (lyrics), pitch and timbre are extracted respectively from the training audio, then decoded back to generate waveform with no timbre transferring. After training stage, timbre conversion can be achieved by simply replacing the source extracted timbre embedding with the target singer's timbre in the conversion stage. However, in order to transfer content, pitch and timbre independently, it is crucial to keep these embedding features accurate and fully disentangled, with no interweaving information in them. Some methods tried to completely disentangle content embedding with timbre and pitch through adversarial training \cite{Nachmani2019UnsupervisedSV,9054199} or VAE \cite{luo2020singing,kameoka2019acvae} modeling, but the converted audio quality and similarity to target singer is harmed. More than enough information is saved for accurate audio reconstruction, which degrades the singing conversion quality and similarity.

Different from above, the PPG-based approach does not extract the features in a self-learning manner. Instead, Phonetic posteriorgrams (PPG) are more robust content embeddings represetations, and it is based on the knowledge and model pre-trained using huge amount of speech recognition data. It has also been proved a reliable and effective approach in both voice conversion  \cite{sun2016phonetic,tian2018average,Polyak2019TTSSS} and singing voice conversion \cite{Chen2019SingingVC,gao2020personalized}. However, compressing waveform to low-resolutional PPG features could cause information loss. In this case, a high quality generator module is a must to compensate the loss and to render high fidelity and naturalness audio. MelGAN neural vocoder proposed in \cite{kumar2019melgan} has demonstrated its strong performance in audio generation and efficiency. In this paper, we propose to link the auto-encoder based singing conversion module with the MelGAN-like neural vocoder module. Adversarial training are also exploited in the system to improve the conversion performance. Our main contributions are as follows:
\begin{itemize}
    \item A generator based on MelGAN is designed to map PPG, F0, speaker embedding to the waveform. Before upsampling input features to the waveform, The CNN-BLSTM based module is employed to extract more efficient sequential information from the PPG and F0 sequence. For more robust and controllable timbre representation, the speaker embedding is further decomposed to the weighted sum of a set of trainable vectors representing different timbre/speaker clusters.
    \item Two discriminators are added in the generator training process. Audio generation discriminator $D_{A}$ trains the generator to achieve better generative performance by classifying the real and generated(fake) audio  at different sample rates. Feature disentanglement discriminator $D_{F}$ tries to encode PPG that is further adversarially trained to remove remaining pitch and timbre information. Experiments show that it can produce more stable and accurate conversion performance.
    \item A two-stage training strategy is adopted to keep the adversarial training process effecient and stable. In proposed training, the generator is firstly trained using multi-resolution STFT (MR-STFT) loss, then the model is further trained with both MR-STFT loss and adversarial loss to make the model more stable.
    \item Subjective tests and experimental analyses are conducted to evaluate different methods. The MOS test results demonstrate the effectiveness of our proposed methods. The proposed system also achieves better performance than the conventional cascade approach and the WaveNet based end-to-end approach in both singing quality and singer similarity. The analyses reveal that the model trained with the proposed training strategy can produce a smoother and sharper formant. We also apply our model in timbre transfer and pitch control to further verify the effectiveness of proposed methods.
\end{itemize}


\section{Related Work}


\subsection{PPG-based Singing Voice Conversion}

Similar with other autoencoder based SVC approaches, this framework maps the features representing content, pitch, timbre of the audio to the waveform via the audio generation module. But the difference is that it exploits other well-trained models or well-implemented tools to get these features with more robust and accurate representation. 

Automatic speech recognition (ASR) aims to accurately recognize all content information from the audio. It is trained with large-scale speech corpus that includes various recording environments, background noise, accent, and pitch, and has achieved high accuracy. So its output probability distribution or hidden vectors can be extracted as the robust content representation, which is called phonetic posteriorgrams (PPG). In the same way, the speaker embedding extracted from the speaker recognition (SR) model is also adopted to represent timbre information, which is highly related to the speaker. Pitch (F0) contains the note of a song and the singing style of a singer, e.g. the trill. In our corpus, there is no serious noise and reverberation, so F0 can be accurately extracted using signal processing based tools from the waveform.

\subsection{Audio Generation}

For the PPG-based SVC, there are usually two main approaches for audio generation, the conventional cascade approach, and the end-to-end approach.

\subsubsection{Cascade approach}

It is composed of two models, an acoustic model and a vocoder. Acoustic model mapps PPG features with low frame rate to low-dimension acoustic features, e.g. Mel spectrograms, then upsampling them to the waveform with a high sample rate via a neural vocoder \cite{van2016wavenet, kalchbrenner2018efficient, prenger2019waveglow}. It can ensure both stable mapping relationship, and high-quality audio generation. So it is widely used in singing audio generation \cite{kim2018korean, gao2020personalized}.

However, the limitation also exists obviously in the framework. Firstly, as the intermediate audio representation, the low-dimensional spectral feature helps model stable mapping relationship, but also removes timbre and pitch related information partially. It leads to the limitation of accurate timbre and pitch reconstruction in the neural vocoder. Secondly, the prediction error in the acoustic model will be passed to the neural vocoder, which then leads to sound quality degradation. To avoid these problems, the end-to-end approach is proposed to realize conversion in one model without any intermediate features.

\subsubsection{End-to-end approach}

It aims to directly map PPG features to the waveform in one model, so the model with both good generative capability and stability is required to ensure stable space mapping and high-quality audio generation. The current mainstream approaches are mostly based on the autoregressive model, e.g. WaveNet \cite{tian2019vocoder}, which has shown great conversion performance in VC. However, its huge model structure and autoregressive samples generation lead to more memory cost, and longer inference time. Moreover, the teacher forcing training algorithm used for autoregressive models needs real samples as inputs, but uses the predicted samples at inference. This gap will result in a problem called exposure bias, which makes it difficult to model a stable and effective mapping relationship from PPG features to the waveform. It may seriously affect the sound quality and intelligibility at inference.

To avoid this problem, it is necessary to find an approach to generate the waveform in parallel. MelGAN \cite{kumar2019melgan} proposes an adversarial training based neural vocoder that can generate high-quality audio samples in parallel. In this paper, we purpose to apply it in SVC. To further exploit the effect of adversarial training, we propose an end-to-end adversarial SVC approach.

\begin{figure*}[htp]
    \centering
    \includegraphics[width=0.80\textwidth]{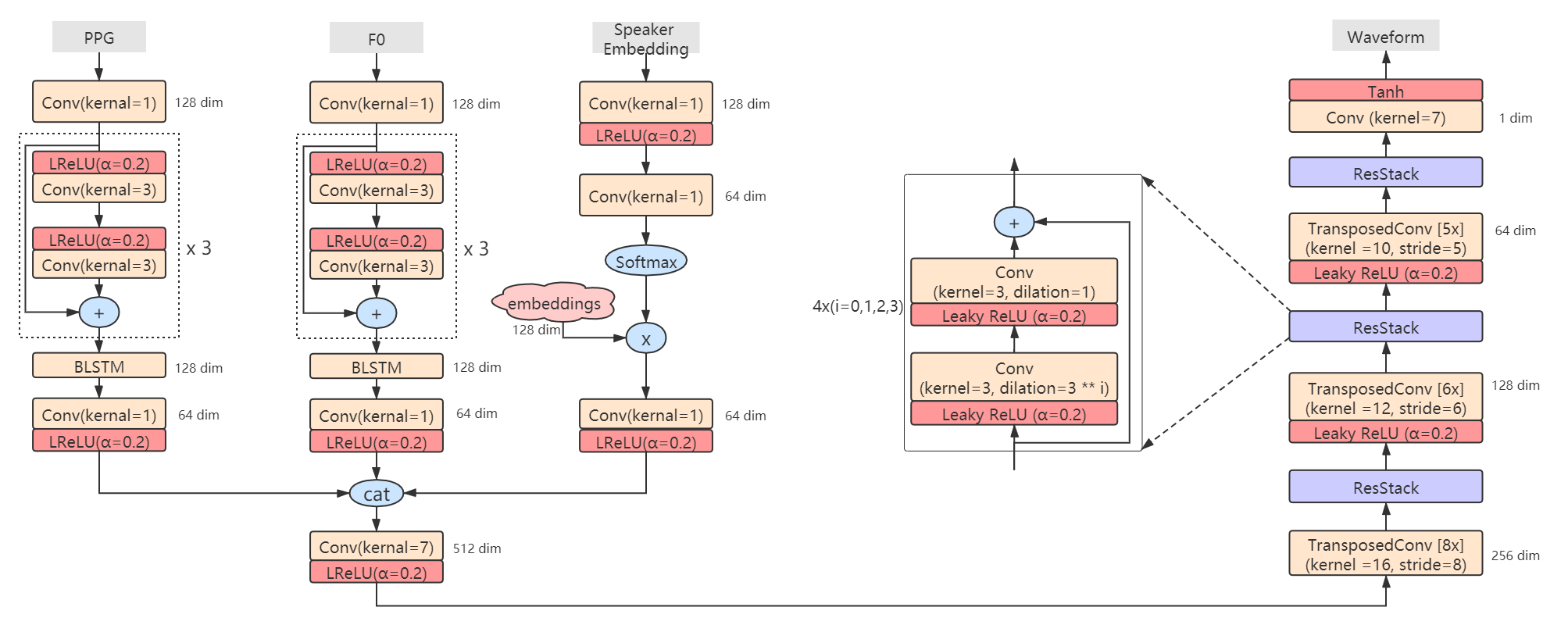}
    \caption{The model structure of the generator in EA-SVC. ``x3" means the three blocks in cascade. The dimension written beside the layer denotes its output channel size.}
    \label{fig:generator}
\end{figure*}

\section{End-to-End Adversarial Singing Voice Conversion}
\label{sec:ea_svc}

In this section, We will introduce this framework in detail, including the generator, two discriminators, and the training algorithm.

\subsection{Generator}

The generator is composed of two parts, an encoder to encode the 218-dim PPG, 1-dim F0 and 128-dim speaker embedding (SE) in parallel, and an upsampler to inversely convert these features to the waveform.

\subsubsection{Encoder}
\label{sec:encoder}

As shown in the left part of figure \ref{fig:generator}, it encodes the three features separately to help better extract sequential information of the feature itself. For PPG, it firstly embeds them into a 128-dim space using a convolutional layer with the kernel size 1. Three residual stacks containing two convolutional layers with the kernel size 3 are used to extract more high-level and contextual information. After it, the BLSTM layer is used to extract the 256-dim vectors with more sequential information. Finally, the convolutional layer with the kernel size 1 compresses them to 64-dim bottleneck features as the content representation. The 1-dim F0 sequence is also processed in the same way.

The timbre representation is extracted in a different way. To perform well in timbre conversion, it is necessary to model a more robust and controllable timbre space. GST-Tacotron \cite{wang2018style} has shown its great controllability in speaking style by controlling the weights on different trainable style tokens. Inspired by it, we propose a speaker embedding decomposition module during encoding it, which is shown as follows:
\begin{gather}
    H_S = W_2(\sigma_L((W_1(S) + b_1)) + b_2 \\
    Y = Softmax(H_S) * E_{1:N}.
\label{eq:se}
\end{gather}
where $W_1$, $W_2$, $b_1$, $b_2$ denote the trainable weight matrixes and biases, $\sigma_L$ denotes “LeakyReLU", and we set $N = 64$ in our work. The speaker embedding $S$ is non-linearly transformed to a 128-dim vector firstly, and then converted to N-dim weight distribution through a Softmax layer. According to the weight vector, the N trainable 128-dim embeddings $E_{1:N}$ are weighted added together, then mapped to a 64-dim vector as the timbre representation. Training these embeddings can be seen as a process of clustering or disentanglement. We can weighted add the N embeddings together  according to the specific proportions to get the expected timbre representation. Notably, we adopt Synthesizer (Dense version) \cite{tay2020synthesizer} to calculate the weight distribution, which can be seen a simplified version of GST.

To concatenate these features together, SE is replicated to the same length with other features. We keep the dimension of these encoded features the same to prevent the model from overfitting to one kind of feature during the training.

\subsubsection{Upsampler}

The structure of the upsampler is similar to MelGAN, which is composed of a stack of transposed convolutional layers followed by residual blocks. A convolutional layer with the kernel size 7 is firstly used to fuse these features to a 512-dim contextual vector. These vectors with the frameshift 10ms are at 240x lower temporal resolution than the waveform with the sample rate 24,000. So three transposed convolutional layers with stride 8, 6, 5 respectively are used to upsample them to the same sample rate. Their kernel sizes are double of the stride to alleviate Checkerboard artifacts \cite{odena2016deconvolution}, which may result in audible high-frequency hissing noise. The deep residual block containing four residual layers with dilation $3^0, 3^1, 3^2, 3^3$ respectively follows each transposed convolutional layer to enhance the model capability. The deeper model structure can help the generator capture more high-level information, which makes the details of the audio more realistic. And the kernel size and wider dilation size can efficiently increase the receptive field, enhance contextual correlation, thus improve the naturalness and smoothness of the generative audio. The final convolutional layer with the kernel size 7 and activation function ``Tanh" generates the waveform normalized between -1 and 1.

There is no noise vector as input in the generator, because diversity is not necessary in our task. We only purpose to improve the fidelity via adversarial training. In addition, to stablize the training process, we apply LeakyReLU with $\alpha = 0.2$ as the default activation function, and weight normalization for all convolutional layers.

\begin{figure}[htp]
    \centering
    \includegraphics[width=0.45\textwidth]{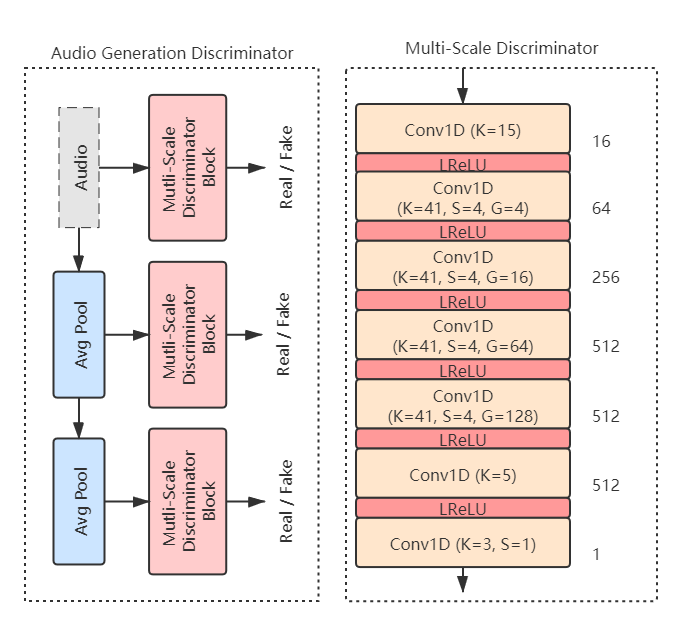}
    \caption{The structure of the discriminator block. $K, S, G$ denotes the kernel size, stride, and groups, respectively. The dimension written at the right of the convolutional layer denotes the output channels of the layer.}
    \label{fig:discriminator}
\end{figure}

\subsection{Audio Generation Discriminator}

The audio generation discriminator $D_{A}$ aims to distinguish if the audio is real or generated. It is employed to adversarially train the generator to improve its generative performance.

\subsubsection{Model architecture}

As shown in figure \ref{fig:discriminator}, it consists of multiple discriminator blocks, which can adversarially train the generator from different aspects. Three multi-scale discriminator blocks are used to deal with the audio at different sample rate, 24000, 12000, and 6000, to capture different features at different frequency range. The audio is downsampled via an average pooling layer with kernal size 4 and stride 2. Each MSD block is a window-based discriminator, which compresses the input audio to small chunks via four strided convolutional layers. Due to the large kernel size in the strided convolutional layer, grouped convolutions are used to keep number of parameters small. In addition, we also apply weight normalization for all layers in the discriminator for stable training process.

\subsubsection{Loss function}

The least-squares GAN version loss is used for $D_{A}$. For each discriminator block $D_{A_k}$, its task is to minimize the loss defined as follows:
\begin{gather}
    L_{adv}(D_{A_k}) = (1 - D_{A_k}(x)) ^ 2 + D_{A_k}(\hat{x}) ^ 2
\label{eq:adv_d}
\end{gather}
where $x$ and $\hat{x}$ denote the real waveform and the generated waveform, respectively.

For the generator, except for the least-squares loss, we also apply feature matching proposed in MelGAN for more effective adversarial training. It minimizes the L1 distance between the feature maps of the real and generated audio in each hidden layer of the discriminator.
\begin{gather}
    L_{FM}(G, D_{A_k}) = \sum^{N_k}_{i=0} \frac{1}{N_k} \left \| D^{(i)}_{A_k}(x) - D^{(i)}_{A_k}(G(c)) \right \|_1 \\
    L_{FM}(G, D_A) = \frac{1}{K} \sum^K_{k=1} L_{FM}(G, D_{A_k})
\label{eq:fm}
\end{gather}
where $c$ denote the input features, $K$ and $N_k$ denote the number of discriminator blocks and the number of the hidden layers (except for the output layer) of the k-th discriminator block, respectively. So the adversarial loss for the generator is defined as:
\begin{gather}
    L_{adv}(G, D_A) = \frac{1}{K} (1 - G(c)) ^ 2 + \lambda_{FM} L_{FM}(G, D_A)
\label{eq:adv_g}
\end{gather}
where $\lambda_{FM}$ is set to 10 in our experiments.

\subsection{Feature Disentanglement Discriminator}

Although PPG represents content information well, the pitch and timbre related information still remain in it partially. Once the model depends on them, it will result in the conflict of information when given different F0 and speaker embedding. The predition will be affected seriously, hence generate inaccurate conversion and noise. So we propose feature disentanglement discriminator $D_{F}$ to remove them from the PPG in an adversarial way.

\subsubsection{Model architecture}

\begin{figure}[htp]
    \centering
    \includegraphics[width=0.40\textwidth]{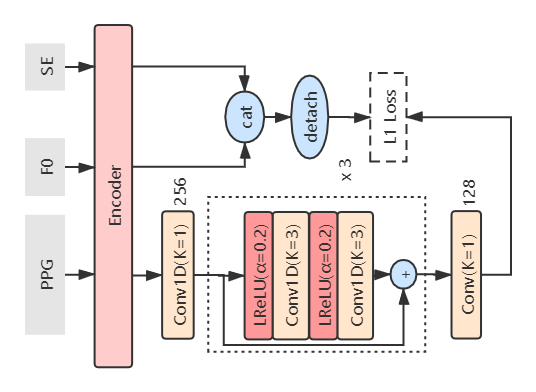}
    \caption{The structure of the feature disentanglement discriminator. When calculating the loss, the gradient is stopped back-propagating to the encoded F0 and SE.}
    \label{fig:FDD}
\end{figure}

As shown in figure \ref{fig:FDD}, $D_{F}$ aims to map the encoded PPG $E_p$ to the vector $E_{fs}$ concatenating the encoded F0 and SE. It adopts the similar structure with the encoder, except for the BLSTM layer. The final convolutional output layer maps the hidden vector to the output with the same dimension with $E_{fs}$.

\subsubsection{Loss function}

L1 loss is used to optimize $D_F$, which is shown as follows:
\begin{gather}
    L_{adv}(D_F) = |\hat{E_{fs}} - E_{fs}|
\end{gather}
where $\hat{E_{fs}}$ denotes the prediction of $D_F$. For the generator $G$, its goal is to make $D_F$ fail to predict $E_{fs}$ from $E_p$. So the loss function for $G$ is:
\begin{gather}
    L_{adv}(D_F) = -1.0 * |\hat{E_{fs}} - E_{fs}|
\end{gather}

\subsection{Training Algorithm}

Weight normalization and advanced model structures are used in adversarial training, but it is still insufficient to stably produce high-fidelity audio. So, in addition to the adversarial loss, we also introduce multi-resolution STFT loss and train the generator using a two-stage strategy. It is an effective method to keep the training process effective and stable \cite{yamamoto2020parallel,yang2020multi}.

\subsubsection{Multi-resolution STFT (MR-STFT) loss}

It computes STFT loss between the real samples $x$ and the generated samples $\hat{x}$ on different resolutions. STFT loss is composed of two loss functions, spectral convergence $L_{sc}$ and log STFT magnitude loss $L_{mag}$ to fit both large and small spectral components \cite{arik2018fast}. It is defined as below:
\begin{gather}
    L_{stft}(x, \hat{x}) = L_{sc}(x, \hat{x}) + L_{mag}(x, \hat{x}) \\
    L_{sc}(x, \hat{x}) = \frac{\left \| |STFT(x)| - |STFT(\hat{x})| \right \|_F}{\left \| |STFT(x)| \right \|_F} \\
    L_{mag}(x, \hat{x}) = \frac{1}{N}\left \| log|STFT(x)| - log|STFT(\hat{x})| \right \|_1
\label{eq:stft}
\end{gather}
where $|STFT(\cdot)|$ and $N$ denote the STFT magnitudes and its dimension respectively, $\left \| \cdot \right \|_F$ and $\left \| \cdot \right \|_1$ denote the Frobenius and $L_1$ norms respectively.

\begin{table}[htp]
\centering
\caption{The analysis parameters of the STFT losses in our work. Hanning window is used in the signal analysis.}
\begin{tabular}{ccc}
\textbf{Frame shift} & \textbf{Frame length} & \textbf{FFT size} \\ \hline
50                   & 240                   & 512               \\ \hline
120                  & 600                   & 1024              \\ \hline
240                  & 1200                  & 2048              \\ \hline
\end{tabular}
\label{tab:stft}
\end{table}

Multi-resolution STFT loss is achieved by averaging multiple STFT losses with different analysis parameters, i.e. frameshift, frame length, and FFT size. It can help generate audio signals with more stable performance in frequency domain. In our work, three STFT losses are used, which are shown in Table \ref{tab:stft}. Finally, the complete loss function for the generator $G$ is:
\begin{gather}
    L_G(G) = L_{stft}(x, G(c)) + L_{adv}(G) \\
    L_{adv}(G) = \lambda_{A} L_{adv}(G, D_A) + \lambda_{F} L_{adv}(G, D_F)
\label{eq:g}
\end{gather}

\subsubsection{Two-stage training strategy}

To make full use of MR-STFT loss, and further stablize the adversarial training process, we propose a two-stage training strategy. In the first stage, we only use $L_{stft}(G, D)$ to train the generator until convergence by setting $\lambda_A = 0$, $\lambda_F = 0$. In the second stage, the generator and the discriminator are trained alternately with $\lambda_A = 4$ and $\lambda_F = 1$. $D_F$ can also be used in the late training process to ensure that the encoder can already provide sufficient information for audio generation.

Stage 1 can help model an accurate and stable mapping relationship between input features and the waveform to ensure stable generated results. It can also be seen as a process of modeling the mean value of the target distribution. Stage 2 exploits adversarial training to replace the averaged value with a more realistic value according to the target distribution. 

\begin{figure}[htp]
    \centering
    \includegraphics[width=0.35\textwidth]{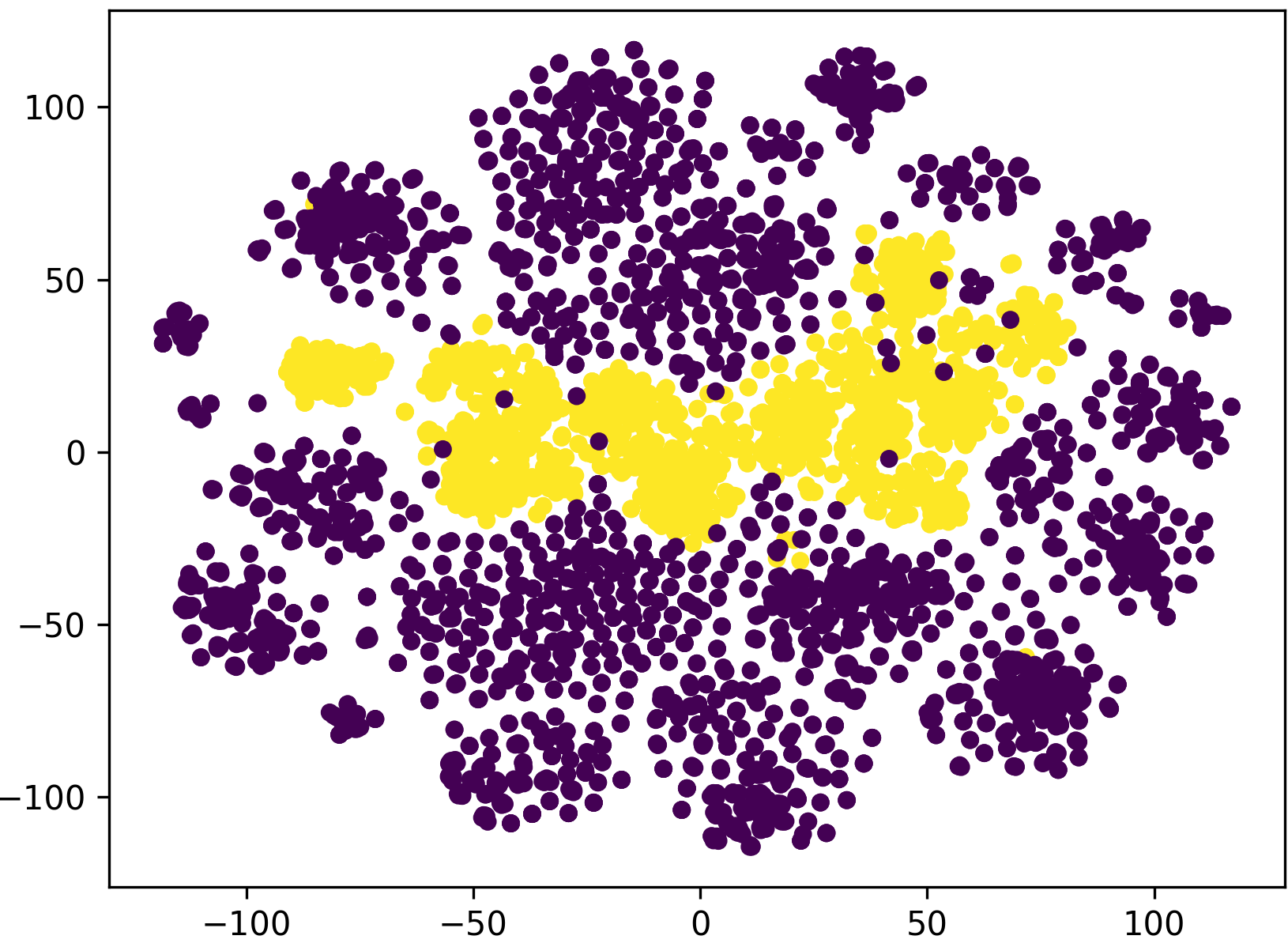}
    \caption{The T-SNE visualization of distribution of speaker embeddings. The yellow and purple dots denote the speaker embeddings from NUS-48E and our dataset, respectively.}
    \label{fig:tsne}
\end{figure}

\section{Experiments}
\label{sec:exp}

\subsection{Experimental Configurations}

\subsubsection{Corpus}

A Mandarin mutli-singer singing corpus is used in our experiments. It contains around 3800 clean singing segments (total 25 hours) of Chinese popular songs filtered from the user data in a sing App. Compared with the open-source English singing corpus NUS-48E \cite{duan2013nus}, it contains more singing content and higher diversity in timbre, which is shown in figure \ref{fig:tsne}. But the scratched dataset also has some problems, e.g. inaccurate pronunciation, uneven singing capability, and various recording environments. It brings a higher requirement for modelling capability.

\subsubsection{Feature extraction}

In our experiments, PPG is the bottleneck feature extracted from a ASR model trained with a 20,000-hour multi-speaker speech corpus. Its frameshift is 10ms. WORLD \cite{morise2016world} is used to extract the F0 with frameshift 10ms. Each singing segment has a speaker embedding, which is a X-Vector extracted from the TDNN model \cite{snyder2018x} trained with a large-scale corpus containing 8800 speakers. We downsample all audios to the sample rate 24,000 for the SVC model training.

\subsubsection{Model configuration}

\begin{figure}[htp]
    \centering
    \includegraphics[width=0.45\textwidth]{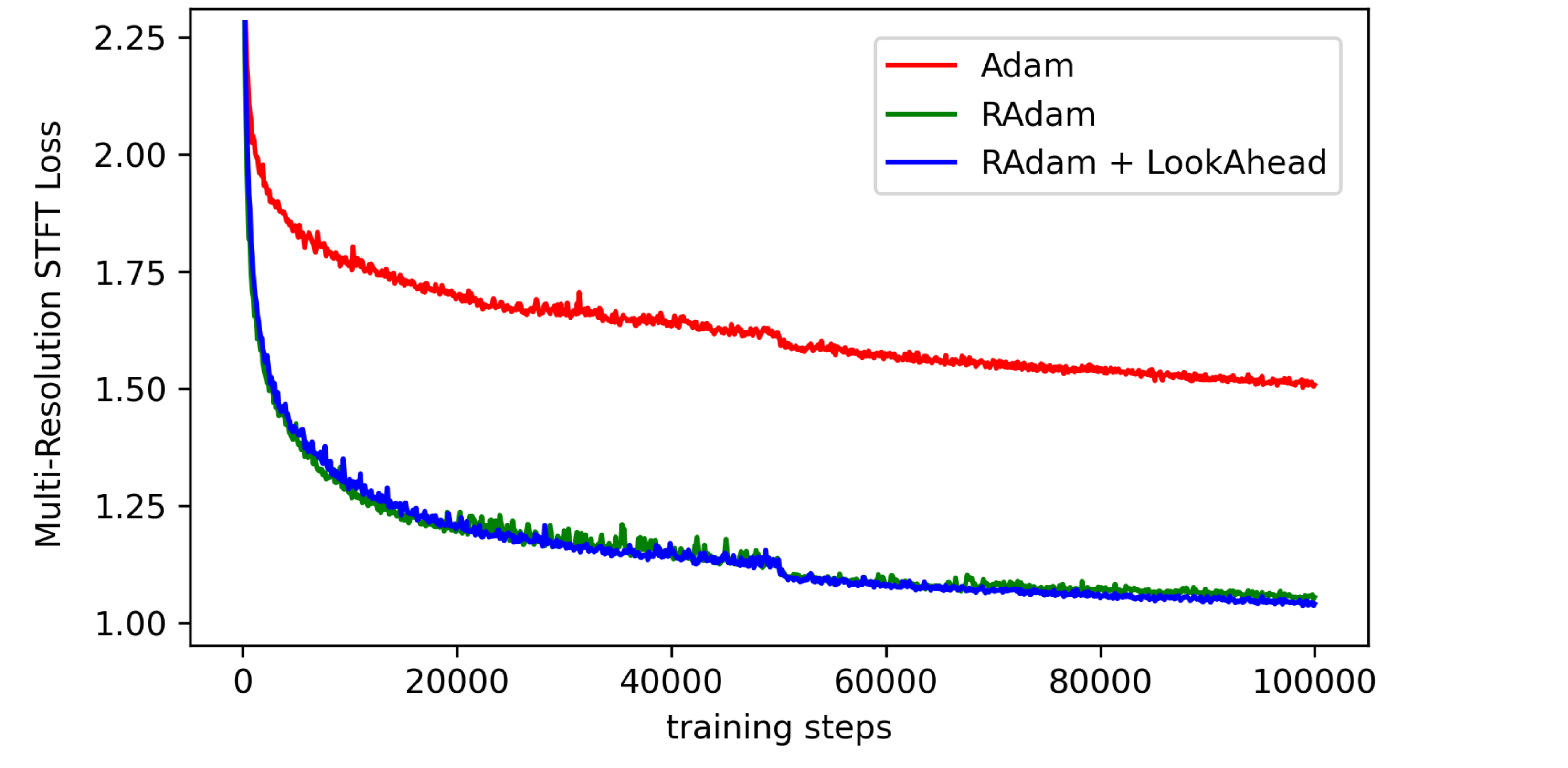}
    \caption{The training loss curves in stage1.}
    \label{fig:loss}
\end{figure}

In our experiments, the models are trained on a single P40 GPU with batch size 32. Each sequence in a batch has 24,000 samples (1 second), which is randomly selected from the audio set. We combine RAdam and LookAhead as the optimizer to speed up and stabilize the whole training process. As shown in Figure \ref{fig:loss}, RAdam \cite{liu2019variance} helps speed up training convergence in the early stage via rectifying the variance of the adaptive learning rate. LookAhead \cite{zhang2019lookahead} can further help stablize the late training process by updating parameters with the slow and fast weights trajectory. In our experiments, the beta of RAdam is set to (0.5, 0.9), the K in LookAhead is set to 5. The learning rate is exponentially decayed based on the training step $t$ using this function:
\begin{equation}
l =
\left\{\begin{matrix}
l_i & t < t_w \\ 
l_{i} * 0.5 ^ {(t / t_w)} & t > t_w \\ 
\end{matrix}\right.
\end{equation}
where $l_i$, $t_w$ denote the initial learning rate and the warmup training steps respectively. The model is trained for 1 million steps totally. In stage 1, we train the generator for 200K steps with $l_i = 10^{-3}, t_w = 50K$. In stage 2, we train the generator and the discrimiantor for 800K steps in adversarial way. We set $l_i = 10^{-4}, t_w = 200K$ for the generator, and a higher learning rate $l_i = 5 * 10^{-4}$ for the discriminator to encourage it to learn faster to capture more distinguishable information. In addition, to stabilize the parameter update of the generator, we also apply the gradient clip with the max 2-norm value of 0.5.

\subsection{Mean Opinion Score (MOS) Test}

We conduct a MOS test to evaluate our proposed approach. There are 4 template songs (2 males, 2 females) in the test set. Each template song is converted to 8 unseen singers (4 males, 4 females), so 32 test cases are included in the set totally. The log-F0 of each song is shifted to be the same as the target singer. The listener needs to give a score for each sample in a test case according to the criterion: 1 = Bad; 2 = Poor; 3 = Fair; 4 = Good; 5 = Excellent. The final score for each model is calculated by averaging the collected results. There are 5 models involved in the evaluation totally.

\subsubsection{Conventional Cascade Approach}

The acoustic model has a similar structure with the encoder in EA-SVC. It inserts two 256-dim BLSTM layers and one output convolutional layer after the encoder. The output, 80-dim Mel-spectrograms, is extracted from the log-magnitude spectrogram with FFT size 2048, frameshift 10ms and frame length 50ms. Its training configuration is similar with stage 1, which trains the model for 200K steps with a batch size of 32, each sequence in a batch has the length of 4 seconds. We also train a MelGAN vocoder with the same configuration for 1M steps to inverse the spectral features to the waveform.

\subsubsection{WaveNet based end-to-end approach}

We implement a WaveNet-based end-to-end SVC model composed of 3 blocks of 10 dilation convolutional layers \footnote{The code links: https://github.com/NVIDIA/nv-wavenet}. The model is trained on the audio with sample rate 16,000, because the model cannot generate intelligible results at sample rate 24,000. It is trained for 800K steps on four P40 GPUs with the batch size of $4 * 16000$ samples totally to achieve sufficient convergence.

\subsubsection{EA-SVC}

To evaluate the effect of the decomposed speaker embedding and $D_F$, we introduce three EA-SVC models in the test, EA-SVC w/o the SE decomposition module and $D_F$, EA-SVC w/o $D_F$, EA-SVC with full configuration. The model w/o decomposition just inputs the speaker embedding to a linear layer as the timbre representation.

\subsubsection{Results}

Table\ref{tab:mos} shows the MOS results of different models. Compare with the conventianal approach (C-SVC) and the WaveNet based approach, EA-SVC achieves the best performance in both singer quality and singer similarity by the MOS of 3.52 and 3.54. C-SVC achieve similar performance in singer similarity, but its quality is not good enough. WaveNet performs badly on our dataset, and causes serious problems in sound quality. We infer that WaveNet is seriously disturbed by exposure bias when trained with the high-diversity dataset. But the model generating waveform in parallel can avoid this problem, hence achieves better performance. \footnote{Samples are availble at \url{https://hhguo.github.io/DemoEASVC}. Our implementation is available at \url{https://github.com/hhguo/EA-SVC}}

\begin{table}[htp]
\centering
\caption{The results of the MOS (95\% CI) test on different models.}
\begin{tabular}{lcc} \toprule
Models & Quality & Similarity \\ \midrule
WaveNet & 2.28 $\pm$ 0.09 & 2.67 $\pm$ 0.14 \\
C-SVC & 3.21 $\pm$ 0.09 & 3.45 $\pm$ 0.10 \\
EA-SVC & 3.52 $\pm$ 0.10  & 3.54 $\pm$ 0.10 \\ \bottomrule
\end{tabular}
\label{tab:mos}
\end{table}

We also use MOS test to evaluate the effect of speaker embedding decomposition and adversarial feature disentanglement. In table \ref{tab:ablation_mos}, the model without decomposition and $D_F$ achieves the worst performance in both quality and similarity. After using decomposition, both scores are improved. It shows that the decomposed speaker embedding can robustly represent timbre information, and performs well in conversion for unseen singers. The introduction of $D_F$ further enhances the singing quality. The unstable problems, like pitch glitches, fuzzy formants, are alleviated during pitch and timbre conversion. It also shows the necessity and effectiveness of the feature disentanglement.

\begin{table}[htp]
\centering
\caption{The results of the MOS (95\% CI) test for ablation studies.}
\begin{tabular}{lcc} \toprule
Models & Quality & Similarity \\ \midrule
w/o Decomposition \& $D_{F}$ & 3.26 $\pm$ 0.11 & 3.35 $\pm$ 0.11 \\
w/o $D_{F}$ & 3.52 $\pm$ 0.12 & 3.55 $\pm$ 0.12 \\ \midrule
EA-SVC & 3.78 $\pm$ 0.11 & 3.60 $\pm$ 0.13 \\ \bottomrule
\end{tabular}
\label{tab:ablation_mos}
\end{table}

\subsection{Analysis}

In this section, we make analysis to evaluate the effect of the proposed modules in singing conversion.

\begin{figure}[htp]
    \centering
    \includegraphics[width=0.45\textwidth]{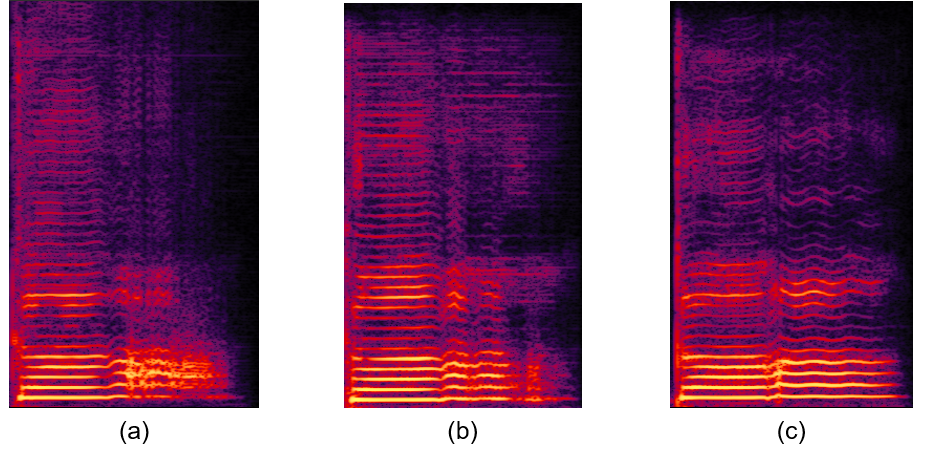}
    \caption{The Mel spectrograms of the audios generated by the models: (a) $M_0$, (b) $M_1$, (c) $M_2$}
    \label{fig:spec}
\end{figure}

\subsubsection{Training strategy}

To show the important of the two-stage training algorithm and MR-STFT loss, we compare the models trained with different training strategies:
\begin{itemize}
    \item $M_0$: Trained in stage2 with $L_{adv}$
    \item $M_1$: Trained in stage2 with $L_{stft}$ and $L_{adv}$
    \item $M_2$: Trained in stage1 + stage2 with $L_{stft}$ and $L_{adv}$
\end{itemize}
Their difference in audio generation can be obviously observed on the spectrogram. As shown in figure \ref{fig:spec}, $M_0$ produces fuzzy formants without the constraint of MR-STFT loss, causing serious problems in intelligibility. Compared with $M_1$, $M_2$ using two-stage training strategy produces much more clear and smooth formants, leading to higher sound quality.

\begin{figure}[htp]
    \centering
    \includegraphics[width=0.45\textwidth]{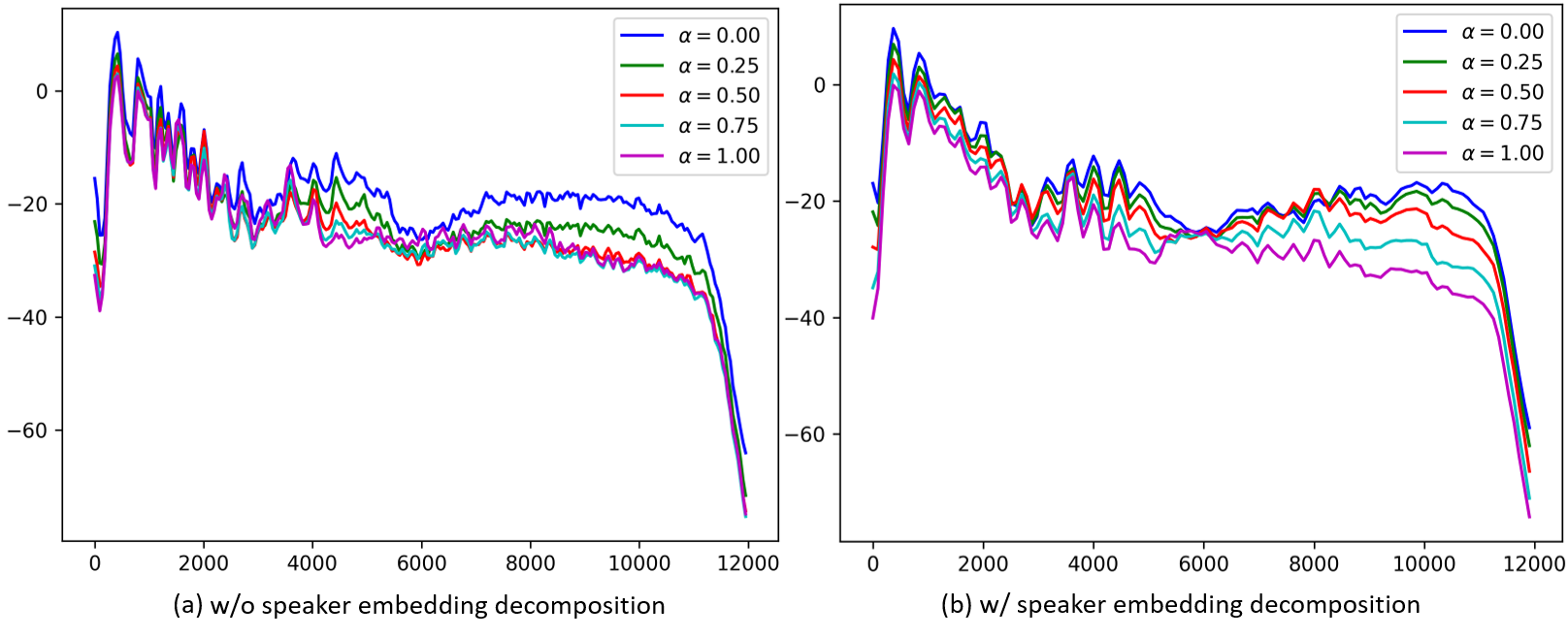}
    \caption{The averaged magnitude spectrogram of the converted audio with different transfer scales.}
    \label{fig:timbre}
\end{figure}

\subsubsection{Timbre Transfer}

In timbre conversion, except for replacing the speaker embedding completely, it is also needed to interpolate between two timbres to achieve partial transfer, which requires a higher controllability of the timbre space. We compare the models w/ or w/o speaker embedding decomposition to investigate their performance in timbre control.

As shown in figure \ref{fig:timbre}, we purpose to gradually transfer the timbre by controlling the scale $\alpha$. For the model w/o decomposition, we weighted add the encoded source SE and target SE. For the model w/ decomposition, we use the same way, but operate on the weight distribution on different timbre clusters. The results show that the right averaged magnitude spectrogram is linearly transferred as $\alpha$ increasing, but the left one does not show this expected trend. It demonstrates that the decomposition can better control the timbre space.

\subsubsection{Pitch control}

\begin{figure}[htp]
    \centering
    \includegraphics[width=0.40\textwidth]{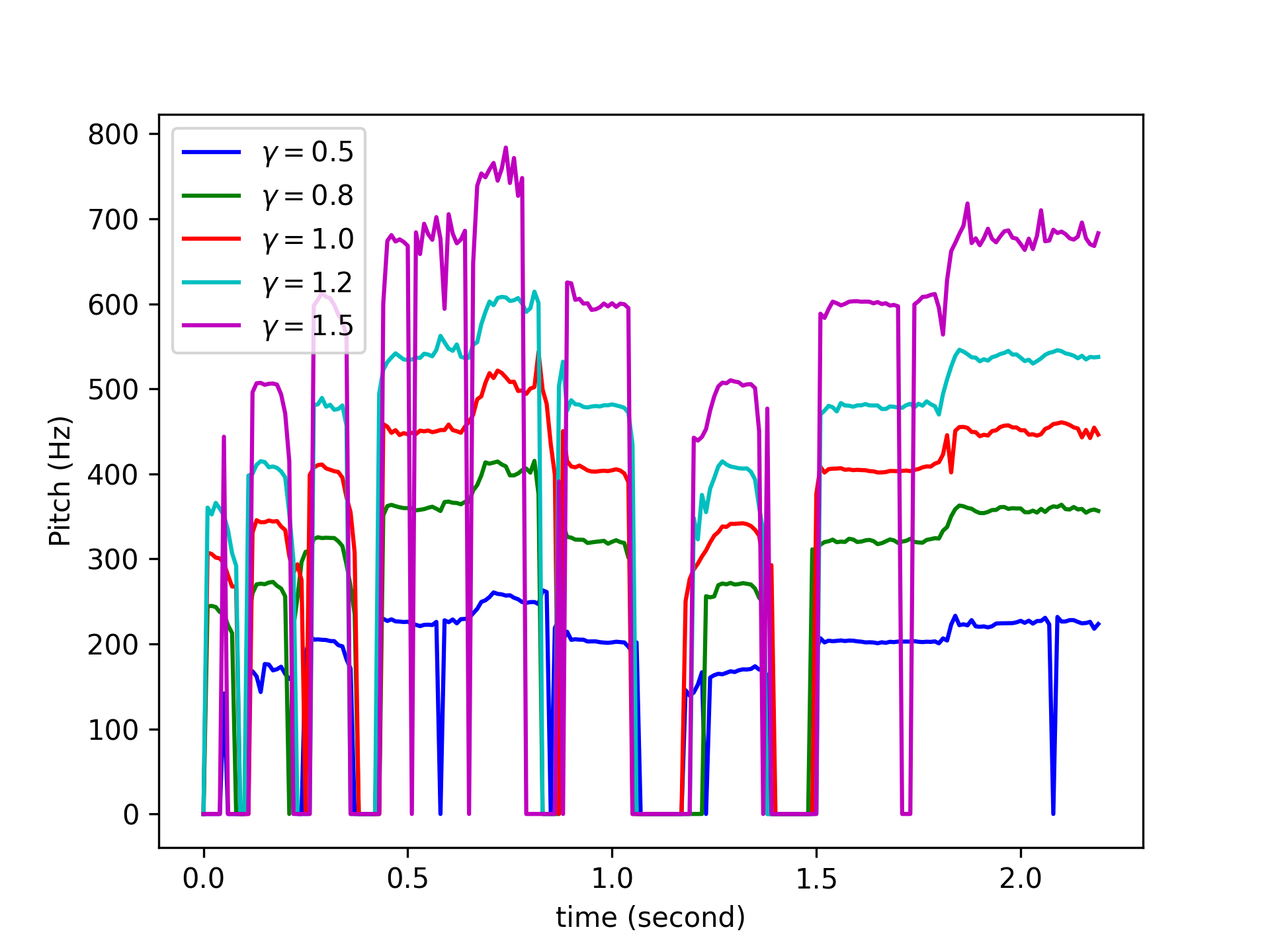}
    \caption{The pitch contours of the same singing segment converted with different scales.}
    \label{fig:pitch_shift}
\end{figure}

We investigate the capability of the model in pitch control. We purpose to globally manipulate the pitch by multiplying F0 with a factor, $\gamma$. For example, the pitch should be an octave (12 keys) higher when $\gamma = 2.0$. We rebuild a song with different $\gamma$, which is shown in figure \ref{fig:pitch_shift}. The results show that the model performs well when controlling $\gamma$ in a normal range, e.g. 0.8 to 1.2, but generated unstable glitches when the pitch is modified too much, e.g. $\gamma=1.5$. We infer that this situation can be alleviated by data augmentation to covers more pitch contours in the train set and further feature disentanglement.

\section{Conclusion}
\label{sec:con}

This paper proposes an adversarial training based end-to-end singing voice conversion approach. In the generator, encoder uses two CNN-BLSTM modules to encode PPG and F0 respectively, and decomposes the speaker embedding into a weight distribution of a group of trainable vectors representing different timbre components. The adversarial training is applied in two aspects, audio generation and feature disentanglement. The multi-scale discriminator is use to adversarially train the generator to produce high-fidelity audio. Another discriminator aiming to map the encoded PPG to the encoded F0 and SE is proposed to remove overlapped information remained in PPG. A two-stage training strategy combining MR-STFT loss and adversarial loss is employed to keep a more stable and effective adversarial training process. We conduct MOS tests to evaluate the proposed methods. The results show that EA-SVC achieves the best performance in both quality and similarity over the conventional cascade approach and the WaveNet based end-to-end apporach. Objective anlysis is also conducted to further demonstrate the effectiveness of our proposed modules.

\bibliography{strings}

\begin{thebibliography}{26}
\providecommand{\natexlab}[1]{#1}
\providecommand{\url}[1]{\texttt{#1}}
\providecommand{\urlprefix}{URL }
\expandafter\ifx\csname urlstyle\endcsname\relax
  \providecommand{\doi}[1]{doi:\discretionary{}{}{}#1}\else
  \providecommand{\doi}{doi:\discretionary{}{}{}\begingroup
  \urlstyle{rm}\Url}\fi

\bibitem[{Ar{\i}k, Jun, and Diamos(2018)}]{arik2018fast}
Ar{\i}k, S.~{\"O}.; Jun, H.; and Diamos, G. 2018.
\newblock Fast spectrogram inversion using multi-head convolutional neural
  networks.
\newblock \emph{IEEE Signal Processing Letters} 26(1): 94--98.

\bibitem[{Chen et~al.(2019)Chen, Chu, Guo, and Xu}]{Chen2019SingingVC}
Chen, X.; Chu, W.; Guo, J.; and Xu, N. 2019.
\newblock Singing Voice Conversion with Non-parallel Data.
\newblock In \emph{MIPR}, 292--296.

\bibitem[{{Deng} et~al.(2020){Deng}, {Yu}, {Lu}, {Weng}, and {Yu}}]{9054199}
{Deng}, C.; {Yu}, C.; {Lu}, H.; {Weng}, C.; and {Yu}, D. 2020.
\newblock Pitchnet: Unsupervised Singing Voice Conversion with Pitch
  Adversarial Network.
\newblock In \emph{ICASSP}, 7749--7753.

\bibitem[{Duan et~al.(2013)Duan, Fang, Li, Sim, and Wang}]{duan2013nus}
Duan, Z.; Fang, H.; Li, B.; Sim, K.~C.; and Wang, Y. 2013.
\newblock The NUS sung and spoken lyrics corpus: A quantitative comparison of
  singing and speech.
\newblock In \emph{2013 Asia-Pacific Signal and Information Processing
  Association Annual Summit and Conference}, 1--9. IEEE.

\bibitem[{Gao et~al.(2020)Gao, Tian, Zhou, Das, and Li}]{gao2020personalized}
Gao, X.; Tian, X.; Zhou, Y.; Das, R.~K.; and Li, H. 2020.
\newblock Personalized Singing Voice Generation Using WaveRNN.
\newblock In \emph{Odyssey 2020 The Speaker and Language Recognition Workshop},
  252--258.

\bibitem[{Kalchbrenner et~al.(2018)Kalchbrenner, Elsen, Simonyan, Noury,
  Casagrande, Lockhart, Stimberg, Oord, Dieleman, and
  Kavukcuoglu}]{kalchbrenner2018efficient}
Kalchbrenner, N.; Elsen, E.; Simonyan, K.; Noury, S.; Casagrande, N.; Lockhart,
  E.; Stimberg, F.; Oord, A. v.~d.; Dieleman, S.; and Kavukcuoglu, K. 2018.
\newblock Efficient neural audio synthesis.
\newblock In \emph{ICML}.

\bibitem[{Kameoka et~al.(2019)Kameoka, Kaneko, Tanaka, and
  Hojo}]{kameoka2019acvae}
Kameoka, H.; Kaneko, T.; Tanaka, K.; and Hojo, N. 2019.
\newblock ACVAE-VC: Non-parallel voice conversion with auxiliary classifier
  variational autoencoder.
\newblock \emph{IEEE/ACM Transactions on Audio, Speech, and Language
  Processing} 27(9): 1432--1443.

\bibitem[{Kim et~al.(2018)Kim, Choi, Park, Hahn, Kim, and Kim}]{kim2018korean}
Kim, J.; Choi, H.; Park, J.; Hahn, M.; Kim, S.; and Kim, J.-J. 2018.
\newblock Korean singing voice synthesis system based on an LSTM recurrent
  neural network.
\newblock In \emph{INTERSPEECH}, 1551--1555.

\bibitem[{Kumar et~al.(2019)Kumar, Kumar, de~Boissiere, Gestin, Teoh, Sotelo,
  de~Br{\'e}bisson, Bengio, and Courville}]{kumar2019melgan}
Kumar, K.; Kumar, R.; de~Boissiere, T.; Gestin, L.; Teoh, W.~Z.; Sotelo, J.;
  de~Br{\'e}bisson, A.; Bengio, Y.; and Courville, A.~C. 2019.
\newblock Melgan: Generative adversarial networks for conditional waveform
  synthesis.
\newblock In \emph{Advances in Neural Information Processing Systems},
  14910--14921.

\bibitem[{Liu et~al.(2019)Liu, Jiang, He, Chen, Liu, Gao, and
  Han}]{liu2019variance}
Liu, L.; Jiang, H.; He, P.; Chen, W.; Liu, X.; Gao, J.; and Han, J. 2019.
\newblock On the variance of the adaptive learning rate and beyond.
\newblock \emph{arXiv preprint arXiv:1908.03265} .

\bibitem[{Luo et~al.(2020)Luo, Hsu, Agres, and Herremans}]{luo2020singing}
Luo, Y.-J.; Hsu, C.-C.; Agres, K.; and Herremans, D. 2020.
\newblock Singing voice conversion with disentangled representations of singer
  and vocal technique using variational autoencoders.
\newblock In \emph{ICASSP}, 3277--3281. IEEE.

\bibitem[{Morise, Yokomori, and Ozawa(2016)}]{morise2016world}
Morise, M.; Yokomori, F.; and Ozawa, K. 2016.
\newblock WORLD: a vocoder-based high-quality speech synthesis system for
  real-time applications.
\newblock \emph{IEICE TRANSACTIONS on Information and Systems} 99(7):
  1877--1884.

\bibitem[{Nachmani and Wolf(2019)}]{Nachmani2019UnsupervisedSV}
Nachmani, E.; and Wolf, L. 2019.
\newblock Unsupervised Singing Voice Conversion.
\newblock In \emph{INTERSPEECH}.

\bibitem[{Odena, Dumoulin, and Olah(2016)}]{odena2016deconvolution}
Odena, A.; Dumoulin, V.; and Olah, C. 2016.
\newblock Deconvolution and checkerboard artifacts.
\newblock \emph{Distill} 1(10): e3.

\bibitem[{Polyak, Wolf, and Taigman(2019)}]{Polyak2019TTSSS}
Polyak, A.; Wolf, L.; and Taigman, Y. 2019.
\newblock TTS Skins: Speaker Conversion via ASR.
\newblock \emph{arXiv preprint arXiv:1904.08983} .

\bibitem[{Prenger, Valle, and Catanzaro(2019)}]{prenger2019waveglow}
Prenger, R.; Valle, R.; and Catanzaro, B. 2019.
\newblock Waveglow: A flow-based generative network for speech synthesis.
\newblock In \emph{ICASSP}, 3617--3621. IEEE.

\bibitem[{Snyder et~al.(2018)Snyder, Garcia-Romero, Sell, Povey, and
  Khudanpur}]{snyder2018x}
Snyder, D.; Garcia-Romero, D.; Sell, G.; Povey, D.; and Khudanpur, S. 2018.
\newblock X-vectors: Robust dnn embeddings for speaker recognition.
\newblock In \emph{ICASSP}, 5329--5333. IEEE.

\bibitem[{Sun et~al.(2016)Sun, Li, Wang, Kang, and Meng}]{sun2016phonetic}
Sun, L.; Li, K.; Wang, H.; Kang, S.; and Meng, H. 2016.
\newblock Phonetic posteriorgrams for many-to-one voice conversion without
  parallel data training.
\newblock In \emph{2016 IEEE International Conference on Multimedia and Expo
  (ICME)}, 1--6. IEEE.

\bibitem[{Tay et~al.(2020)Tay, Bahri, Metzler, Juan, Zhao, and
  Zheng}]{tay2020synthesizer}
Tay, Y.; Bahri, D.; Metzler, D.; Juan, D.-C.; Zhao, Z.; and Zheng, C. 2020.
\newblock Synthesizer: Rethinking Self-Attention in Transformer Models.
\newblock \emph{arXiv preprint arXiv:2005.00743} .

\bibitem[{Tian, Chng, and Li(2019)}]{tian2019vocoder}
Tian, X.; Chng, E.~S.; and Li, H. 2019.
\newblock A vocoder-free WaveNet voice conversion with non-parallel data.
\newblock \emph{arXiv preprint arXiv:1902.03705} .

\bibitem[{Tian et~al.(2018)Tian, Wang, Xu, Chng, and Li}]{tian2018average}
Tian, X.; Wang, J.; Xu, H.; Chng, E.~S.; and Li, H. 2018.
\newblock Average Modeling Approach to Voice Conversion with Non-Parallel Data.
\newblock In \emph{Odyssey}, volume 2018, 227--232.

\bibitem[{Van Den~Oord et~al.(2016)Van Den~Oord, Dieleman, Zen, Simonyan,
  Vinyals, Graves, Kalchbrenner, Senior, and Kavukcuoglu}]{van2016wavenet}
Van Den~Oord, A.; Dieleman, S.; Zen, H.; Simonyan, K.; Vinyals, O.; Graves, A.;
  Kalchbrenner, N.; Senior, A.~W.; and Kavukcuoglu, K. 2016.
\newblock {WaveNet}: A generative model for raw audio.
\newblock In \emph{SSW}, 125.

\bibitem[{Wang et~al.(2018)Wang, Stanton, Zhang, Skerry-Ryan, Battenberg, Shor,
  Xiao, Ren, Jia, and Saurous}]{wang2018style}
Wang, Y.; Stanton, D.; Zhang, Y.; Skerry-Ryan, R.; Battenberg, E.; Shor, J.;
  Xiao, Y.; Ren, F.; Jia, Y.; and Saurous, R.~A. 2018.
\newblock Style tokens: Unsupervised style modeling, control and transfer in
  end-to-end speech synthesis.
\newblock \emph{arXiv preprint arXiv:1803.09017} .

\bibitem[{Yamamoto, Song, and Kim(2020)}]{yamamoto2020parallel}
Yamamoto, R.; Song, E.; and Kim, J.-M. 2020.
\newblock Parallel WaveGAN: A fast waveform generation model based on
  generative adversarial networks with multi-resolution spectrogram.
\newblock In \emph{ICASSP}, 6199--6203. IEEE.

\bibitem[{Yang et~al.(2020)Yang, Yang, Liu, Fang, Chen, and
  Xie}]{yang2020multi}
Yang, G.; Yang, S.; Liu, K.; Fang, P.; Chen, W.; and Xie, L. 2020.
\newblock Multi-band MelGAN: Faster Waveform Generation for High-Quality
  Text-to-Speech.
\newblock \emph{arXiv preprint arXiv:2005.05106} .

\bibitem[{Zhang et~al.(2019)Zhang, Lucas, Ba, and Hinton}]{zhang2019lookahead}
Zhang, M.; Lucas, J.; Ba, J.; and Hinton, G.~E. 2019.
\newblock Lookahead optimizer: k steps forward, 1 step back.
\newblock In \emph{Advances in Neural Information Processing Systems},
  9597--9608.

\end{thebibliography}

\end{document}